\documentclass[aps,prb,twocolumn,groupedaddress,floatfix,showpacs]{revtex4-2}

\usepackage{graphicx}
\usepackage{amsmath}
\usepackage{subcaption}
\usepackage{color}
\newcommand{\parl}{\parallel}
\newcommand{\beq}{\begin{equation}}
\newcommand{\bearr}{\begin{eqnarray}}
\newcommand{\eeq}{\end{equation}}
\newcommand{\earr}{\end{eqnarray}}
\begin{document}


\title{Optical properties of excitons in CdSe nanoplatelets and disks:
real density matrix approach}


\author{David Ziemkiewicz}
\email{david.ziemkiewicz@utp.edu.pl} ,
\author{Gerard Czajkowski}
\author{Sylwia Zieli\'{n}ska-Raczy\'{n}ska}
 \affiliation{Institute of
Mathematics and Physics, Technical University of Bydgoszcz,
\\ Al. Prof. S. Kaliskiego 7, 85-789 Bydgoszcz, Poland}


\date{\today}

\begin{abstract}  
We show how to calculate the optical functions of a nanoplatelet, taking into account the effect
of a dielectric confinement on  excitonic states. Real density matrix approach is employed to obtain analytical and semi-analytical relations for the absorption coefficient, the exciton resonance energy and binding energy of nanoplatelets and nanodisks. The impact of plate geometry (thickness, area) on the spectrum is discussed and the results are compared with the available experimental data. 
\end{abstract}



\maketitle
\section{Introduction}

The quantum size effects in semiconductor nanocrystals, first pointed out in 1981 \cite{Ekimov} have unlocked plethora of research topics for semiconductor nanosystems \cite{Yu}. 
The synthesise of nearly monodispersive semiconductor nanocrystallites, such as cadmium selenide (CdSe) \cite{Murray} opened the way to  producing nanosystems of various dimmensions; zero dimensional quantum dots where electrons and holes are confined in three dimensions,   one dimensional (1D) nanorodes with 2D-confined carriers,  and between them are   nanopaletels  (NPLs) of a large lateral size but of only a few molecular layers of thickness, with 1D-confined electrons and holes.
  
 Compared with zero-dimensional quantum dots and one-dimensional nanorods, two-dimension nanoplatels exhibit many unique optical properties. 
Colloidal two-dimensional semiconductor nanoplatelets, which are atomically flat,  exhibit quantum confinent only in one dimension, which results in an electronic structure that is significantly different compared to that of other quantum-confinent nanomaterials. The lateral size of these systems  can range from a few to tens of nanometers, while in the transverse direction the thickness of NPLs can reach only a few atomic layers, which is smaller then the exciton Bohr radius. The exciton confinement is very strong in the direction perpendicular to the plane of NPLs, but the in-plane motion is free.
  Cadmium selenide  NPLs, first fabricated in 2006 \cite{Joo} have become  important examples of a two-dimensional colloidal nanosystem, with large exciton binding energy, strong quantum  confinement and  huge oscillator strength, which allows for a high tunability of their optical properties \cite{Tessier}; they  exhibit strong and narrow emission lines at both cryogenic and room temperatures \cite{Diroll}. The strong Coulomb interaction leads to exciton binding energy reaching hundreds of meV (the bulk  binding energy is only 15 meV); it is remarkable quality of CdSe NPLs, which strongly  affects their optical properties. 
  
 The light-matter interactions in 2D CdSe NPLs offer several unique advantages in optoelectronic and photonic applications.
A high exciton binding energy combined with a specific shape and a confined potential, which limits exciton degrees of  freedom  in NPLs, lead to a strong exciton-photon interaction in these systems and enables  applications of that kind of systems in photonic devices, light-emitting diodes, sensing and light harvesting \cite{Yu,Dutta} and references therein.  The remarkable  thinness of these materials provides  unique opportunities for engineering the excitonic properties.

It should be mentioned that nanoplatelets are specific quantum wells (QW), in which   electrons and  holes are confined in a layer of one type of
semiconductors by an impenetrable barrier of a different
semiconductor. In the CdSe NPLs considered below  the confinement
is of electrostatic origin and is caused by a large dielectric
mismatch between the semiconductor (here CdSe) and its
environment. Despite this difference, electrons and holes interact
by a screened Coulomb potential.  The bound electron-hole pairs created by a propagating electromagnetic wave are
named excitons and they determine  the optical properties of the medium.
 Optical properties of excitons  strongly depend on  systems size and on  shapes of  a nanostructure. In a bulk semicnductors  excitons resemble hydrogen-like quasiparticles with free center-of- mass motion in all directions, but in quantum dots electrons and holes are confined and so are  excitons.
 As in generic QWs, the optical
properties of NPLs are dominated by excitons. So the major part of
research on NPLs is concentrated on the calculations of exciton
characteristics, such as exciton binding energy, confinement
eigenfunctions, and eigenvalues. 
  Recently several groups have measured the exciton binding energy of CdSe NPLs \cite{Benchamekh,Shornikova,Zelewski}.

But the sole knowledge of excitonic
binding energy is not sufficient to describe, interpret and
explain the observed optical spectra. As was pointed by 
Shornikova\emph{ et al.} \cite{Shornikova},  any approaches
which allow one to predict the exciton parameters and
their impact on the optical properties are desirable.
In the presented work we hope to satisfy this expectation.

The recent growth of interest in such systems encourages us to present a method, which gives a simple analytical expression for optical functions, taking into account confinement and dielectric potentials and any excitonic states, which enables one to obtain theoretical spectra. A majority of authors, describing theoretically the optical properties of excitons in NPLs, are using perturbation calculus, where unperturbed eigenfunctions, describing the carriers motion in the $z$-direction, are the (finite- or infinite) 1-dimensional quantum well functions. The binding energy is then calculated with Coulomb e-h interaction potential considered as perturbation \cite{Bagda,Brumberg}. A numerical attempt to calculate electronic properties of CdSe NPLs spectra has also been taken by Benchamekh \textit{et al} \cite{Benchamekh} who applied an advanced tight-binding model. In the present work we propose a different confinement potential, resulting directly from the dielectric confinement. This potential allows for  an analytical solution of the Schr\"{o}dinger equation for electron and hole, describing their motion in the $z$-(confinement) direction. 

 We propose a simple confinement potential, which allows an analytical
solution of the Schr\"{o}dinger equation for electron and hole,
describing their motion in the $z$- (confinement) direction. We
obtain both eigenfunctions and eigenvalues, which enables to
determine the exciton binding energy. Since the solution of the
Schr\"{o}dinger equation for in plane relative motion is well
known, we can estimate the total binding energy. Moreover,  we present the theoretical method, which allows one to calculate  optical properties (i.e., positions of resonances and absorption spectra)  of NPLs and nanodiscs depending on numbers of monolayers (i.e. thickness of NPLs) .  
This method called real density matrix approach (RDMA), which takes into account the effects of an anisotropic dispersion and coherence of the electron and the hole with radiation field, allows one to obtain analytical expressions for the  susceptibillity.

The paper is organized as follows. In Sec. II we recall the basic
equations of the used approach (RDMA), adapting them to the case
of CdSe NPLs. Sec. III is devoted to specific calulations for nanoplatelets and nanodisks with finite
lateral extension. Sec. IV contains theoretical results  and comparison with experimental data of the binding energy for systems of different shapes, while in sec. V  absorption spectra for NPLs and disks are shown are discussed.  The last section presents the concluding remarks.

\section{Real density matrix approach}\label{sec_RDMA}
In this section we briefly review the basic principles of the real density matrix method in the context of linear optical properties of excitons in semiconductors.  
One of the adventages of RDMA is its formulation 
 in the real space. The method gives a direct relation
between the density matrices and the relevant observables, which
allows for a  straight  comparison between  experimental and theoretical
results. The RDMA has been effectively
applied to the optical excitation spectra of semiconductor bulk
crystals, semiconductor superlattices and some low-dimensional
structure (see Ref.\cite{Rivista} for a selection
of earlier references). After the discovery of the so-called
Rydberg excitons, see, for example,
Refs.\cite{Kazimierczuk,AssmannBayer_2020} and references therein,
RDMA has been successfully applied to describe their optical
properties, both linear and nonlinear, see
Refs.\cite{Ziemkiewicz_24,Morin}.\\\\
 The approach starts in the framework of second
quantization. The lowest level consists of two-point density
matrices, which describe the inter-band transitions between the
valence and the conduction band and the intra-band transitions in
the indicated bands. Those quantities are important, since there
are related to direct measurable quantities as polarization and
carrier densities.\\
The real density
matrix approach  takes into account the following contributions:
\newline
a) the electron-hole interaction,
\newline
b) the dipole interaction between the electron-hole pairs and the
electromagnetic field,
\newline
c) the particle-surface interaction,
\newline
d) effects of external fields.
\newline \\
The basic equations 
are obtained in the following way.  
We consider a semiconductor in the real space representation,
characterized by  a number of valence and conduction bands.
Electrons at site $j$ in the conduction band are described by
fermion operators $\hat{c}^{c\dag}_{j} (\hat{c}^c_{{j}})$ which
correspond to   creation (annihilation) operators. Similarly,
operators $\hat{d}^{v\dag}_{{j}} (\hat{d}^v_{{j}})$ are creation
(annihilation) operators for holes in valence bands of a  $j$ state.
In the case of direct interband transitions  Hamilton 
in our model consists of three parts
\begin{equation}\label{2quanhamiltonian}
H=H_0+H_{em}+{H}_C.
\end{equation}
\noindent The term $ {H}_0$ describes one-particle Bloch states in
conduction and valence bands and also  the intra-band
transport processes, the operator $H_{em}$ is an
interaction with the electromagnetic field. The physical
quantities which are most relevant for the optical properties can
be expressed in terms of mean values of the following pair
operators
\begin{eqnarray}
&&\hbox{excitonic transition density amplitude}\nonumber\\
&&Y_{12}^{\alpha\,b}=\langle\hat{Y}_{12}^{\alpha\,b}\rangle
=\langle\hat{d}^{\alpha}_1\,\hat{c}^b_2\rangle,\nonumber\\
&&\hbox{electron density}\quad
C_{12}^{ab}=\langle\hat{C}^{ab}_{12}\rangle=\langle{\hat{c}^{a\dag}}_1\hat{c}^b_2\rangle,\\
&&\hbox{hole density}\quad
D_{12}^{\alpha\,\beta}=\langle{\hat{D}^{\alpha\,\beta}}_{12}\rangle=\langle{\hat{d}^{\alpha\dag}}_1\hat{d}^\beta_2\rangle,\nonumber
\end{eqnarray}
\noindent where the indices $a,b,..$ 
and $\alpha,\beta,...$ stand for conduction and valence bands, respectively. The excitonic
transition density $Y^{\alpha\,b}_{12}$ contributes to the polarization by the following term
\begin{equation}\label{polcrystal}
P= 2\,\hbox{Re}\,\left(\int_{r=r_1-r_2}d^3r\,
\sum_{cv}\,M_{21}^{cv*}Y_{12}^{vc}\right),
\end{equation}
where $M^{cv}_{21}$ is the interband dipole matrix element
and the diagonal elements (the matrices $C$ and $D$) correspond to
the densities of electrons
\begin{equation}\label{densityelectrons}
\rho_e=\left.-e\sum_c\,C^{cc}_{12}\right|_{r_1=r_2},
\end{equation}
\noindent and holes
\begin{equation}\label{densityholes}
\rho_h=\left.e\sum_v\,D^{vv}_{12}\right|_{r_1=r_2}.
\end{equation}
\noindent These matrices are submatrices of the
following density matrix
\begin{equation}\label{matrix}
\underline{\underline{\hat{\rho}}}=\begin{pmatrix}
C_{cc'}&Y^*_{vc} \cr Y_{vc}&1-D_{vv'}\end{pmatrix}.
\end{equation}
\noindent The dynamics of the two-point functions $Y,C,D$ is part
of the hierarchy of reduced density matrices and is obtained from
the Heisenberg equations
\begin{equation}\label{Liouville}
i\hbar\partial_t\underline{\underline{\hat{\rho}}}
=[{H},\underline{\underline{\hat{\rho}}}]+i\hbar\partial_t\underline{\underline{{\hat{\rho}}}}_{\rm
irrev},
\end{equation}
\noindent where the term
$\partial_t \underline{\underline{{\hat{\rho}}}}_{\rm irrev}$  describes
the dissipation and radiation decay processes, which  are due to all
dephasing phenomena. In many practical calculations all
irreversible processes are described in terms of two dephasing
times $T_1, T_2$ which are taken as phenomenological constants and
satisfy the following equation
\begin{eqnarray}\label{Liouv2}
&&\left.\frac{\partial\underline{\underline{\hat{\rho}}}}{\partial
t}\right|_{irrev}
=\\
&&\quad=-\begin{pmatrix}\frac{1}{T_1}[C(t)-C^{(0)}]&\frac{1}{T_2}[Y^*(t)-Y^{*(0)}]\cr
\frac{1}{T_2}[Y(t)-Y^{(0)}]&\frac{1}{T_1}[D(t)-D^{(0)}]
\end{pmatrix},\nonumber
\end{eqnarray}
where the states with superscript $^{(0)}$ denote the steady state
solutions.\\ Equations (\ref{Liouville}) then become a closed set of
differential equations (usually called constitutive  or
band-edge equations) for $Y, C, D$ after the following operations \cite{HuhnStahl84,Stahl84}:\\
\begin{enumerate}
\item setting up the Heisenberg equations of motion for the pair
operators, \item  applying anticommutation rules for the Fermion
operators $\hat{c}_j^{c\dag}, \hat{c}_j^c, \hat{d}_j^{v\dag}$ and
$\hat{d}_j^v$ to bring all operator products into normal
order,\item  going over to expectation values,\item using an
interpolation procedure to obtain  a continuum dependence on the
position variables (for example, Ref.\cite{StB87}).
\end{enumerate}
 As a result we obtain the constitutive equations for inter-band transition density amplitudes , which for any couple of bands are of the following form
\begin{eqnarray}\label{genconstitut}
&&-i\hbar
\partial_tY_{12}+H_{eh}Y_{12}\\
&&={M_0}\left(E\delta_{12}-E_1C_{12}-E_2D_{21}\right)-
i\hbar\left(\frac{\partial Y_{12}}{\partial t}\right)_{\rm
irrev}.\nonumber
\end{eqnarray}
\noindent For intra-band transitions (time dependence of the
population of the band states) one has
\begin{eqnarray}
&&\label{CD}-i\hbar
\partial_tC_{12}+H_{ee}C_{12}\\
&&=-{M_0}\left(E_1Y_{12}-E_2Y^*_{21}\right)-
i\hbar\left(\frac{\partial C_{12}}{\partial t}\right)_{\rm irrev},\nonumber\\
&&\label{DV}
-i\hbar\partial_tD_{12}+H_{hh}D_{12}\\
&&=-{M_0}\left(Y_{21}E_1-Y^*_{12}E_2\right)-
i\hbar\left(\frac{\partial D_{12}}{\partial t}\right)_{\rm
irrev}\nonumber.
\end{eqnarray}
\noindent  The numerical subscripts are abbreviations for
coordinates in the sense $Y_{12}=Y({\bf r}_1,{\bf r}_2)$. Hamiltonian $H$  describes propagation in
$(\textbf{r}_1,\textbf{r}_2)$ space and consists of three parts   
$$H=H_{ee}+H_{hh}+H_{eh},$$  
which are given by
\begin{eqnarray}\label{Heh1}
&&H_{eh}=E_g-V_{12}+\frac{1}{2m_h}\left(p_1-eA_1\right)^2+\frac{1}{2m_e}\left(p_2+eA_2\right)^2\nonumber\\
&&+e\left(\Phi^h_1-\Phi^e_2\right),
\end{eqnarray}
the electron Hamiltonian
\begin{equation}
H_{ee}=\frac{1}{2m_e}\left[(p_2+eA_2)^2-(p_1-eA_1)^2\right]+e\left(\Phi^e_1-\Phi^e_2\right),
\end{equation}
the hole Hamiltonian
\begin{equation}\label{Hhh1}
H_{hh}=\frac{1}{2m_h}\left[(p_2-eA_2)^2-(p_1+eA_1)^2\right]-e\left(\Phi^h_1-\Phi^h_2,\right)
\end{equation}
where $E_g$ denotes the gap energy, $m_e, m_h$ are the effective
masses of electrons and holes, respectively, $V_{12}$ is the
statically screened Coulomb potential, $M_0$ interband transition
element 
 integrated over the real space, $A_j$ is
the vector potential of the Maxwell field at position ${\bf r}_j$,
which may include an external magnetic field, $\Phi^{e/h}_j$ a
scalar, external or electromagnetically induced potential acting
on electrons (or holes) at position ${\bf r}_j$. $E_j$ denotes
the electric field of the radiation at the point ${\bf r}_j$.  We
neglect for the moment the vectorial and tensorial indices, and
use the common notation for the momentum operators: $p_1=
-i\hbar\nabla_1$ etc. The above equations must be solved
simultaneously with the Maxwell field equations
\begin{equation}\label{Maxwell}
-c^2\epsilon_0\hbox{\boldmath$\nabla$}\times\hbox{\boldmath$\nabla$}\times\textbf{E}-\epsilon_0\epsilon_b\ddot{\textbf{E}}=\ddot{\textbf{P}},
\end{equation}
where the polarization \textbf{P} is given by Eq. (\ref{polcrystal}) and
$\epsilon_b$ is the bulk dielectric constant. When the effects of
confinement are considered, one makes use of the appropriate boundary
conditions for $\textbf{E}, Y, C$ and $D$. 
$$\,$$
In the weak field limit and for a multiband semiconductor the set
of equations  (\ref{genconstitut}-\ref{DV}) for the linear case
 reduces to a set of
linearized constitutive equations, which are the inter-band
equations (\ref{genconstitut}), where we put $C=D=0$ on the
right-hand side. The resulting equations for the excitonic
amplitudes$Y_{12}^{\alpha b}$ of the electron-hole pair of
coordinates $\textbf{r}_1=\textbf{r}_h$ and
$\textbf{r}_2=\textbf{r}_e$ between any pair of bands $\alpha$ and
$b$ have the form
\begin{equation}\label{Lin1}
-i(\hbar\partial_t+\Gamma_{\alpha b})Y_{12}^{\alpha b}+H_{eh\alpha
b}Y_{12}^{\alpha b}=\textbf{M}_{\alpha b}\textbf{E},\end{equation}
where $\Gamma_{\alpha b}=\hbar/T_2^{\alpha b}$  is a
phenomenological damping coefficient, see Eq. (\ref{Liouv2}). The
two-band Hamiltonian $H_{eh\alpha b}$ with energy gap $E_{g\alpha
b}$ for any pair of bands reads
\begin{eqnarray}\label{Ham1}
H_{eh\alpha b}&=&E_{g\alpha
b}+\frac{\textbf{p}^2_{h\alpha}}{2m_\alpha}+\frac{\textbf{p}^2_{eb}}{2m_b}+V_{eh}(1,2)\nonumber\\
&+&V_h(1)+V_e(2),\end{eqnarray} with electron and hole kinetic
energy operators, $m_\alpha$ and $m_b$ being the band effective
masses, $V_{eh}$ describes the electron-hole attraction and $V_e$,$V_h$ denote the confinement
potentials of the electron and the hole, respectively. The total polarization of the
medium (\ref{polcrystal}) is related to the excitonic amplitudes
by
\begin{equation}\label{polarization}
\textbf{P}(\textbf{R})=2\sum\limits_{\alpha,..,a,...}\hbox{Re}\,\int
d^3r \textbf{M}_{cv}(\textbf{r})
Y^{cv}(\textbf{R},\textbf{r}),\end{equation}
 where $\textbf{r}=\textbf{r}_e-\textbf{r}_h$ is the relative
coordinate, and $\textbf{R}$ the electron-hole pair center-of-mass
coordinate, and the summation includes all allowed excitonic
transitions between the valence 
and
conduction bands. 

\section{Basic equations}
\subsection{Nanoplatelets}
 For CdSe based NPLs we have to 
consider both,  heavy(H) and light(L) hole excitons. For the
optical transitions between ($\alpha=H,L$) valence bands and the
conduction band ($b=C$) we get two constitutive equations for
the excitonic amplitudes $Y_{12}^{HC}=Y_H(r_e,r_h),\,
Y^{LC}_{12}=Y_L(r_e,r_h)$
\begin{eqnarray}\label{HL}
&&-i\hbar\partial_tY_H-i\Gamma_HY_H+H_{ehH}Y_H=\textbf{M}_H(\textbf{r})\,\textbf{E}(\textbf{R}),\nonumber\\
&&-i\hbar\partial_tY_L-i\Gamma_LY_L+H_{ehL}Y_L=\textbf{M}_L(\textbf{r})\,\textbf{E}(\textbf{R}),\end{eqnarray}
where $\textbf{M}_{H,L}(\textbf{r})$ are transition dipole
densities. The operators $H_{ehH,L}$ has the form
\begin{eqnarray}\label{HL2}
&&H_{ehH,L}=E_{gH,L}+\frac{p_{ez}^2}{2m_{ez}}+V_{e}(z_e)\\
&&+\frac{p_{hz}^2}{2m_{hzH,L}}+V_h(z_h)
+\frac{\textbf{P}_{\parallel}^2}{2M_{\parallel
H,L}}+\frac{\textbf{p}_\parallel^2}{2\mu_{\parallel H,L}}\nonumber\\
&&+V_{eh}(z_e-z_h,\rho),\nonumber\end{eqnarray}
where we have separated the center-of-mass coordinate $\textbf{R}_\parallel$ and
the related momentum $\textbf{P}_\parallel$ from the relative
coordinate $\rho$ on the plane $(x,y)$ and the related momentum
$\textbf{p}_\parallel$. $M_{\parallel H,L}=m_{h\parallel H,L}+m_{e\parallel}$ is the total in-plane exciton mass. Using Hamiltonian (\ref{HL2}) we obtain the
constitutive equations in the form
\begin{eqnarray}
&&(H_{ehH}-\hbar\omega-i\Gamma_H)Y_H(\rho,z_e,z_h)\\
&&=\textbf{M}_H(\hbox{\boldmath$\rho$},z_e,z_h)\textbf{E}(\textbf{R}_\parallel,Z).\nonumber\end{eqnarray}
An analogous equation  can be written  for the light hole exciton amplitude
$Y_L$.\\For a state $N$ we will seek solutions in the form
\begin{equation}\label{expansionH}
Y_H(\hbox{\boldmath$\rho$},z_e,z_h)=\sum_{j,m,N}c_{jm
NH}\psi_{jm}(\hbox{\boldmath$\rho$})\psi_{eN}(z_e)\psi_{hNH}(z_h),\end{equation}
where $\psi_{jm}(\hbox{\boldmath$\rho$})$ are the eigenfunctions
of the so-called 2-dimensional hydrogen atom, resulting from the
equation
\begin{equation}
\left(\frac{\textbf{p}_\parallel^2}{2\mu_{\parallel
H}}-\frac{e^2}{4\pi\epsilon_0\epsilon_b\rho}\right)\psi_{jm}(\rho)=E_{jm}\psi_{jm}(\rho),\end{equation}
and have the form
\begin{eqnarray}\label{eigenfunctions}
&&\psi_{jm}(\hbox{\boldmath$\rho$})=\frac{1}{a_{e\parallel^*}}\frac{e^{im\phi}}{\sqrt{2\pi}}\nonumber\\
&&\times
e^{-2\lambda\rho/a_{e\parallel^*}}(4\lambda\rho)^m\,4\lambda^{3/2}\frac{1}{(2m)!}\frac{[(j+2m)!]^{1/2}}{[j!]^{1/2}}\nonumber\\
&&\times M\left(-j,2\vert m\vert+1,4\lambda\rho\right)\nonumber\\
&&=R_{jm}(\rho)\frac{e^{im\phi}}{\sqrt{2\pi}},\\
&&\lambda=\frac{1}{1+2(j+\vert m\vert)},\nonumber
\end{eqnarray}
with the confluent hypergeometric function $M(a,b;x)$, defined as
an infinite series \beq\label{series}
M(a,b;x)=1+\frac{a}{1!b}x+\frac{a(a+1)}{2!(b+1)}x^2+\ldots.\eeq
The eigenvalues, corresponding to the eigenfunctions
(\ref{eigenfunctions}), are given by
\begin{equation}\label{eigenvalues}
\frac{E_{jm}}{R_\parallel^*}=\varepsilon_{jm}=-\frac{4}{[1+2(j+\vert
m\vert)]^2},
\end{equation}
where $R_\parallel^*$ is the Rydberg energy for the in-plane data and $a_{e\parallel}^*$ is the excitonic Bohr radius defined as
\begin{equation}
a_{\parallel,v}^*=\frac{1}{\mu_{\parallel,v}}\epsilon_1 a^*_B, \qquad v=H,L.
\end{equation}
For clarity of the details of subsequent calculations  it is sound to mention here that  functions $\psi_{e,h N}$ are eigenfunctions of the
one-dimensional Schr\"{o}dinger equations of the type
\begin{equation}\label{S1}
\frac{p_{z}^2}{2m_{z}}\psi_{e,h N}(z)+V(z)\psi_{e,h N}(z)=E_z\psi_{e,h N}(z),
\end{equation}
with a given potential $V(z)$.
For further calculations we have to define the form of the
confinement potentials $V_{e,h}(z_{e,h})$. The interaction between
charged carriers (electrons and holes) with their is described
a repulsion potential. In the case of a great mismatch between  the dielectric constant of external
media $\epsilon_2$ ("$\,$out$\,$") and the dielectric
constant of internal media $\epsilon_1$ ("$\,$int$\,$") ($\epsilon_2<<\epsilon_1$), the
"$\,$mirror force$\,$" is repulsive. This also leads  to a repulsive
potential that pushes the charge $e$ away from the surface and this force can be described by the repulsive potential, which is
inversely proportional to distance from the surface and diverges
at least at one of the interfaces. Considering a particle confined
in a slab of thickness $L$, with the surfaces at $z=\pm L/2$, one  can
solve the Schr\"{o}dinger equation with a specific one-dimensional
model potential, which, with regard to the above description, can be
taken in the form \beq\label{potential}
V_{e,h}(z)=\frac{\gamma_{e,h}}{(L/2)-z}.\eeq The coefficient $\gamma$
is proportional to dielectric coefficients
\cite{Landau, Caicedo}
$$\gamma \propto
\frac{\epsilon_1-\epsilon_2}{\epsilon_1(\epsilon_1+\epsilon_2)}.$$
Apart from the confinement potential (\ref{potential}), both
carriers (electron and hole) interact also by a Coulomb potential, which leads to the following form of eq. (27)
\begin{eqnarray}\label{1}
&&-\frac{\hbar^2}{2m_z}\frac{d^2}{dz^2}\psi_{e,h N}(z)\nonumber\\
&&+\frac{\gamma\,R_z^*\,a_z^*}{(L/2)-z}\psi_{e,h N}(z)=E_z\psi_{e,h N}(z),
\end{eqnarray}
where $m_z$ denotes the effective mass of the considered quasi
particle (electron or hole) in the $z$-direction, $R_z^*,
a_z^*$ are the corresponding effective Rydberg energies, and Bohr
radii defined as
 \begin{eqnarray}\label{definitions}
&&\label{Reh}R_{e,hz}^*=\frac{m_{e,h z}\,e^4}{2(4\pi\epsilon_0\epsilon_1)^2\hbar^2}\\
&&=\left(\frac{m_{e,hz}}{m_0}\right)\frac{1}{\epsilon_1^2}R^*_B,\nonumber\\
&&\label{aeh}a_{e,hz}^*=\frac{\hbar^2(4\pi\epsilon_0\epsilon_1)}{m_{e,hz}\,e^2}=\left(\frac{m_0}{m_{e,hz}}\right)\epsilon_b\,a^*_B,
\end{eqnarray}
where $m_0$ is the free electron mass, $R^*_B=13600 $ meV and $a^*_B=0.0529 $ nm are the
hydrogen Rydberg energy, and Rydberg radius, respectively. The detailed calculations of 
eigenfunctions and eigenvalues related to Eq. (\ref{1}) is presented
in Appendix A ((\ref{ehz3}) and (\ref{eigendiel})). They
correspond to the lowest excitonic confinement state, so we omit
the index $N$ in (\ref{expansionH}). Making use of the functions
(\ref{eigenfunctions}) and (\ref{eigendiel}),  we calculate the
expansion coefficients $c_{jmv}$,$v=H,L$,
\begin{equation}\label{cj}
c_{jmv}=\frac{\langle M^*_v(\hbox{\boldmath$\rho$}, z_e,z_h)\vert
\psi_e(z_e)\psi_h(z_h)
\psi_{jmv}(\hbox{\boldmath$\rho$})\rangle}{E_{g}-\hbar\omega-i\Gamma_v+E_{conf
v}-E_{bjmv}},
\end{equation} 
where $E_{conf v}$ is given in Eq.
(\ref{conf24}), and $E_{bjmv}$ is the so-called binding energy
\begin{equation}\label{binding}
E_{bjmv}=2\int\limits_0^\infty\rho\,d\rho\int\limits_0^{{\mathcal
L}_e}d\zeta_e\int\limits_0^{{\mathcal
L}_h}d\zeta_h\frac{R_{jm}^2(\rho)\psi_e^2(z_e)\psi_h^2(z_h)}{\sqrt{\rho^2+(z_e-z_h)^2}}.
\end{equation}
With the use of  the above coefficients we obtain the
amplitudes $Y_H$ and $Y_L$, which determine the NPL polarization
by Eq. (\ref{polarization}), which in turn enables one to obtain the
NPL susceptibility and then, the optical functions.

\subsection{Disks with lateral confinement}\label{disklateral}
It seems that aside from nanoplatelets, nanodisks can also be  an important systems in the context of  accomplishment and experiments. They are two-dimensional structures of cylindrical symmetry and lateral confinement which may be more suitable for preparation and practical applications. Brumberg \emph{et al}.\cite{Brumberg} measured the absorption of CdSe disks. The authors have considered rectangular
plates with vertical dimensions typical of the mostly considered
CdSe NPLs; systems of such a shape need
more complicated theoretical description, which we will be presented below. First, the quadratic  shape will be replaced
by a cylindrical disk with a radius 
appropriate to  given dimension. The lateral confinement potential for
electrons and holes in such a case is given by the expression
\begin{eqnarray}\label{boundarydisk}
 &
&V_{e,h}(\rho_{e,h})=\left\{ \begin{array}{ll}
0\quad \mbox{for}\quad \rho_{e,h}\leq R,\\
\infty\quad \mbox{for}\quad \rho_{e,h}>R,
\end{array}\right.
\end{eqnarray}
 Due to the fact that one of
the carriers (here the hole) has an effective mass much larger
than the other we consider the motion of an electron in
the potential (\ref{boundarydisk}) supplemented with the Coulomb
interaction with the hole, located, in the mean, in the disk
center \cite{ABC}. We will use both negative and positive total
energies for the lateral motion.\\\\ For the case of negative
energies, the respective eigenfunction for the electron motion has
the form\cite{ABC} \bearr\label{eigenfdisk}
\psi_{jm}(\xi,\phi)&=& C\,\xi^{\vert m\vert}e^{-\xi/2}\\
&&\times M\left(m+\frac{1}{2}-\eta,2\vert
m\vert+1;\xi\right)\frac{e^{im\phi}}{\sqrt{2\pi}},\nonumber\earr
where $j$ and $m$ are the principal and magnetic quantum numbers
of the excitonic state, $\rho=\rho_e$,
$$\eta=\frac{2}{\kappa},\quad \xi=\kappa\,\rho, \quad
\kappa^2=-4\frac{2m_{e\parallel}}{\hbar^2}a^{*2}_{e\parallel}E=2/\sqrt{\epsilon}.$$
The quantities $\kappa,\; \rho$, and $\xi$ are dimensionless
(defined in the parameters related to the electron). The
eigenfunction, due to the no escape boundary condition, satisfies the equation
\begin{equation}\label{zeros}
\psi_{jm}(\alpha {\mathcal R},\phi)=0,\quad {\mathcal
R}=\frac{R}{a^*_{e\parallel}},
\end{equation}
giving the eigenenergies $E_{jm}, m=0,1,\ldots,\;j=0,1,\ldots$. In
the linear approximation the first zero will be found from the
equation \beq\label{fitsrt_zerojm}
1+\frac{2m+2j+1-2\eta_{jm}}{2m+1}\left(\frac{2}{\eta_{jm}}{\mathcal
R}\right)=0,\eeq from which one gets the energy $E_{jm}$ 
\begin{eqnarray}\label{lambdamj}
\eta_{jm}&=&\frac{(2m+2j+1){\mathcal R}}{2{\mathcal
R}-2m-1}=\frac{2m+2j+1}{2}+\nu_{jm},\nonumber\\
\nu_{jm}&=&\frac{(2m+2j+1)(2m+1)}{2[2{\mathcal R}-2m-1]},\\
&&E_{jm}=\frac{1}{\eta_{jm}^2}R^*_{e\parallel}.\nonumber
\end{eqnarray}
Here, we only consider the transverse motion of electron, which does not depend on H-L splitting. Therefore, the index $v$ is omitted in $E_{jm}$.)

With regard to the definitions
(\ref{lambdamj}) one can see, that in the limit ${\mathcal R}\to\infty$
the  energies fulfill  the expression (26), 
 characterizing the energy of m-states for the so-called
2-dimensional hydrogen atom ($m=0$ for $S$-states and\\ $m=1$ for
$P$-states etc.). From Eqs. (\ref{lambdamj}) we obtain the values
of critical radii ${\mathcal R}_{cr}$ \bearr\label{rcrit}
&&2{\mathcal R}_{cr}-2m-1=0,\nonumber\\
&&{\mathcal R}_{cr}=\frac{2m+1}{2}.\earr Eqs. (\ref{lambdamj}) are
applicable for ${\mathcal R}>{\mathcal R}_{cr}$.
\\
In
the case of positive energy,  the electron eigenfunction has
the form\cite{ABC}
\bearr\label{eigpositive}  \psi_{jm}(\rho)&=&C
e^{-\xi/2}\xi^{\vert m\vert}\nonumber\\
&\times&\,M\left(\vert m\vert
+\frac{1}{2}-i\eta, 2\vert m\vert+1;\xi\right),
\earr with a normalization
constant $C$.  The eigenenergy can be calculated from  the condition
\beq\label{BCpositive} \hbox{Re}\,\psi_{jm}({\mathcal R})=0.\eeq 
With the help of above eqations we can experess the disk eigenfunctions
in a similar way as for NPLs in Sec. III A. We solve the constitutive
equations (\ref{HL}), with the Hamiltonian
\begin{eqnarray}\label{HL2disk}
&&H_{ehH}=E_{gH}+\frac{p_{ez}^2}{2m_{ez}}+V_{e}(z_e)\\
&&+\frac{p_{hz}^2}{2m_{hzH}}+V_h(z_h)
+\frac{\textbf{P}_{\parallel}^2}{2M_{\parallel
H}}+\frac{\textbf{p}_\parallel^2}{2\mu_{\parallel H}}\nonumber\\
&&-\frac{e^2}{4\pi\epsilon_0\epsilon_b\rho_e}+V_{e}(\rho_{e}),\nonumber
\end{eqnarray}
where we use the dielectric confinement potentials
(\ref{potential}), and $V_e(\rho_e)$ is defined by Eq.
(\ref{boundarydisk}). Then we look for the solutions in the form
(\ref{expansionH}), using the confinement functions
(\ref{eigendiel}), and the in-plane eigenfunctions
$\psi_{jm}(\rho_e)$ (\ref{eigenfdisk}). The expansion coefficients
(\ref{cj}) have now the form
\begin{equation}\label{cjdisk}
c_{jm}=\frac{\langle M(\hbox{\boldmath$\rho$}, z_e,z_h)\vert
\psi_e(z_e)\psi_h(z_h)
\psi_{jm}(\hbox{\boldmath$\rho$})\rangle}{E_{g}-\hbar\omega-i\Gamma+E_{conf
v}-E_{jm}},
\end{equation} where $E_{conf v}$ is given in Eq.
(\ref{conf24}), and $E_{jm}$ are given in Eq. (\ref{lambdamj}),
with appropriate values for the heavy- and light-hole excitons.
It should be noted, that these energies contain effects both from the lateral confinement, and the e-h Coulomb interaction.

The next steps are analogous to those described in Sec. II. With such
calculated  excitonic amplitudes   the polarization and then the disk susceptibility can be determined.
\section{Calculation of the binding energy}
\begin{figure}[ht!]
\centering
\includegraphics[width=.9\linewidth]{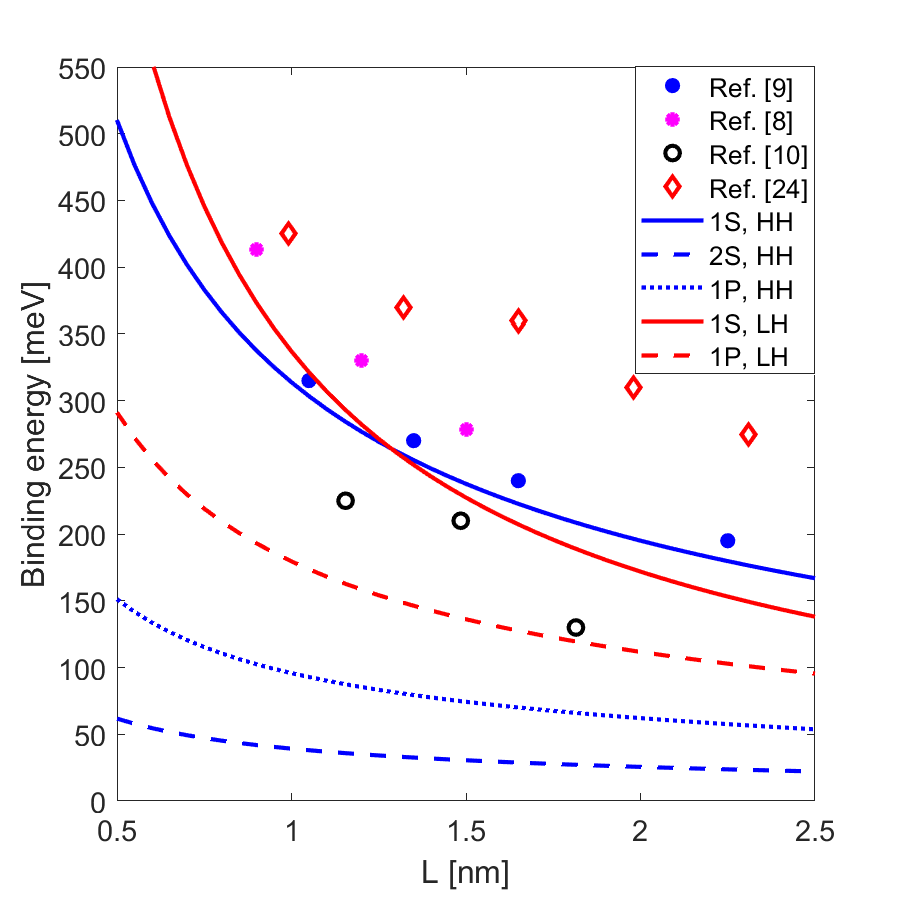}
\caption{Comparison of calculated binding energy with literature data \cite{Shornikova,Benchamekh,Zelewski,Ji}.}\label{fig:bind}
\end{figure}

Since the hole masses (both heavy- and light) are considerably
larger than the electron masses, one can neglect the hole contribution in Eq.
(\ref{binding}), obtaining 
\begin{equation}\label{binding2}
\vert
E_{bjmv}\vert=2\int\limits_0^\infty\rho\,d\rho\int\limits_0^{{\mathcal L}_e}dz_e\,\frac{R_{jm}^2(\rho)\,\psi^2_{e}(z_e)}{\sqrt{\rho^2+p_vz_e^2}}a^*_{ez}R^*_{\parl v}
\end{equation}
where the factor $p_v=\mu_{\parl v}/m_{ez}$ is due to different scaling of $\rho$ and $z_e$. 
With Eq. (\ref{binding2}), we obtain the binding energy shown on Fig. \ref{fig:bind}.
For the $1S$ excitonic state, the heavy hole binding energy is consistent with the data presented in Refs. \cite{Shornikova,Benchamekh}, while different computation models  assumed effective masses result in a similar general tendency, but different value of binding energy \cite{Benchamekh,Ji}.  Significant differences between various models are discussed in \cite{Shor_2018}. 
Our calculations, based on a different confinement potential model, seem to result in a value that is roughly an average of the available literature data.
 Interestingly, both light- and heavy hole energies are very similar, which results in a significant overlap of these states in the absorption spectrum, as  will be shown later. The $1P$ state energies are considerably smaller. Note that the binding energy is strongly dependent on the effective masses; for the details of their fits to the available literature data, see Appendix B.

In the case of nanodisks, one can estimate the binding energy as a function of the disk area. The energies (\ref{lambdamj}) contain contributions from both the confinement energy and the
binding energy. To extract the binding energy, we can use the
perturbation calculus. The unperturbed eigenfunctions are the
solutions of Eq. (\ref{binding2}) and have the form 
\begin{equation}\label{bind_d_1}
\psi_{nm}(\rho)=N_{mn} J_{m}\left(\frac{\rho}{\mathcal R}j_{mn}\right),
\end{equation}
where $N_{mn}$ are the corresponding normalization factors, and $j_{mn}$ the zeros of the Bessel
functions $J_{m}$, $\mathcal R$ is defined in Eq. (\ref{zeros}). With the
eigenfunctions (\ref{bind_d_1}), we can apply the formula (\ref{binding2}), where now
\begin{equation}\label{bind_d_2}
\vert E_{bmn}\vert=2\int\limits_0^{\mathcal R}\int\limits_0^{{\mathcal L}_e}\rho\,d\rho
dz_e\frac{\psi_{mn}^2(\rho)\psi^2_{e}(z_e)}{\sqrt{\rho^2+p_e^2 z_e^2}}\,a^*_{ez}R^*_{\parl e},
\end{equation}
where
\begin{eqnarray}\label{bind_d_3}
&&p_e=\frac{m_{e\parl}}{m_{ez}},\quad R^*_{\parl e}=\frac{m_{e\parl}
R^*_B}{\epsilon_1^2},
\end{eqnarray}

\begin{figure}[ht!]
\centering
\includegraphics[width=.9\linewidth]{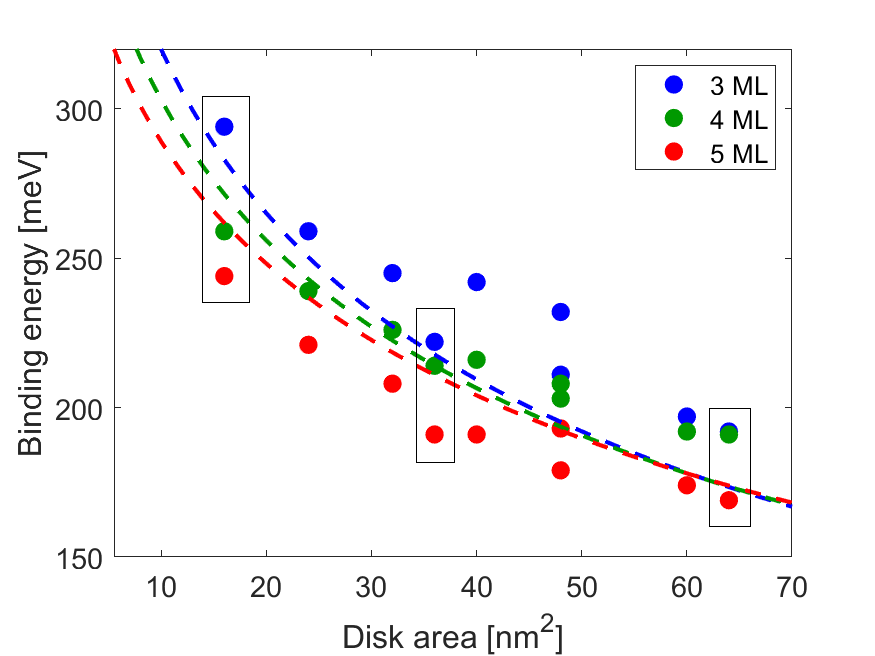}
\caption{Comparison of calculated binding energy  with  \cite{Brumberg}. Boxes mark the data for square platelets.}\label{fig:bind2}
\end{figure}

The binding energies calculated with Eq. (\ref{bind_d_1}) are shown on Fig. \ref{fig:bind2}.
Our calculations result in a fairly good match to the data in Ref. \cite{Brumberg}, correctly predicting the general tendencies and confirming that modeling a square plate as a disk is a valid approximation. Interestingly, even for non-square plates the model works fairly well. Again, it should be stressed that the results depend heavily on the assumed model of the effective masses. 

As a next step, we can calculate the energy of S and P excitons to estimate the S-P splitting. 
\begin{figure}[ht!]
\centering
\includegraphics[width=.9\linewidth]{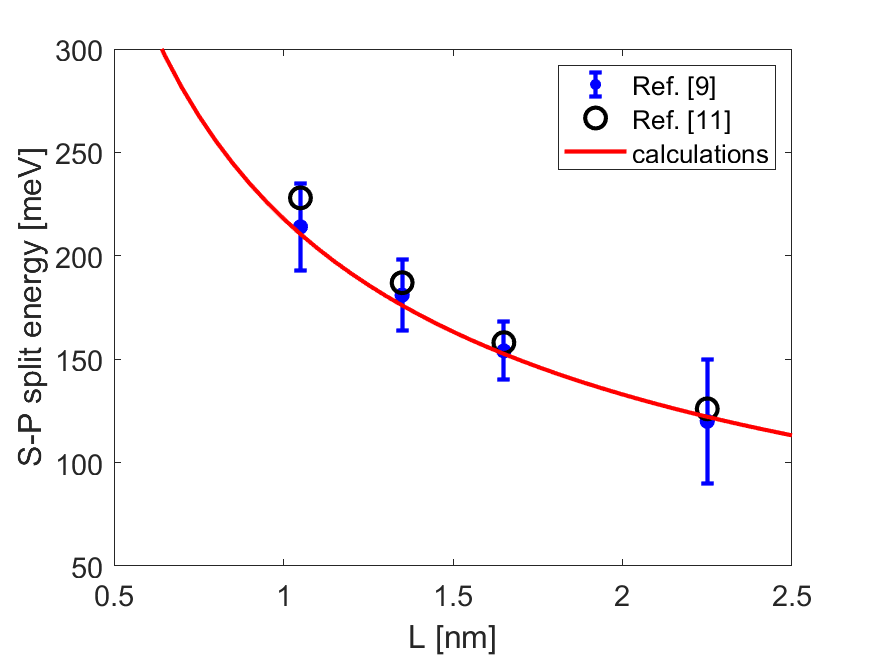}
\caption{Comparison of calculated S-P splitting energy with Refs. \cite{Shornikova,Bagda} }\label{fig:split}
\end{figure}
A comparison of the calculated splitting with the experimental data from \cite{Shornikova} and model from \cite{Bagda} is shown on Fig. \ref{fig:split}. We get an excellent agreement with available measurements \cite{Shornikova,Bagda}.

\section{Calculation of absorption spectrum}
In the considered NPLs widths the typical wavelength of the input
electromagnetic wave is much larger than the NPLs width, so we can
use the long wave approximation. For further calculations we
need to define the dipole density function \textbf{M}. The
transition dipole density $M(\hbox{\boldmath$\rho$})$ should have
the same symmetry properties as the solution of the corresponding
Schr\"{o}dinger equation. Therefore, in the case of 2-dimensional
systems,  we apply the dipole density $M_{m}(\rho,\phi)$  in the
form \bearr\label{Mm}
&&M_{mv}(\rho,\phi;z_e,z_h)=\\
&&=\frac{M_{0mv}\rho^m}{(m+1)!\rho_{0v}^{m+2}}e^{-\rho/\rho_{0v}}\frac{e^{im\phi}}{\sqrt{2\pi}}\delta(z_e-z_h),\nonumber\earr
where $\rho_{0v}=r_{0v}/a^*_{\parl v}$ are the so-called coherence
radii, defined as \beq \rho_{0v}=\sqrt{\frac{R^*_{\parl
v}}{E_{gv}}}.\eeq
\\
 The above equation also indicates, that the integrated
dipole strength $M_{0mv}$ and its relation to the
longitudinal-transversal energy will depend on the quantum number
$m$. Since we will consider both $S$ and $P$ excitons, the
expressions for $m=0$ and $m=1$ have to be used. The dipole matrix elements for $m=0$ have the form  \cite{Rivista} \bearr\label{m0}
M_{00H}^2&=&\frac{\pi}{2}\epsilon_0\epsilon_ba^{*3}\Delta_{LT},\nonumber\\
M_{00L}^2&=&\frac{1}{3}M_{00H}^2. \earr
For $m=1$ the coefficient
$M_{01H}$ and the coherence radius $r_{0H}$ are related to each other by
the longitudinal-transversal energy $\Delta_{LT}$ described by the following relation \cite{ABC}
\bearr\label{M01}
&&(M_{01H}\rho_{0H})^2=\frac{4}{3}\frac{\hbar^2}{2\mu_{\parl H}}\epsilon_0\epsilon_ba^*\frac{\Delta_{LT}}{R_H^*}\,e^{-4\rho_{0H}}\nonumber\\
&&=\frac{4}{3}\epsilon_0\epsilon_ba^{*3}\Delta_{LT}, \earr $a^*$
being the bulk exciton radius.
 With
the above definitions we have all elements to calculate the
mean NPL susceptibility. For the
normal incidence both the electric field of the wave propagating
in the NPL and the polarization (\ref{polarization}) depend on the
center-of-mass coordinate $Z$.
 The mean NPL susceptibility can be written in the form
\begin{equation}\label{meanchi}
\overline{\chi}=\frac{1}{L}\int_{-L/2}^{L/2}\frac{P(Z)}{\epsilon_0E(Z)}
dz.
\end{equation}
Making use of the equations (\ref{expansionH}),
(\ref{polarization}), (\ref{m0}), (\ref{M01}), and taking into
account the relevant resonances, we obtain the susceptibility of
the considered CdSe NPL
\bearr\label{chi1} &&\overline{\chi}(\omega,L)=
A_{ehH}\epsilon_1\Delta_{LT}\left(\frac{\mu_{\parl\,H}}{\mu_{\parl\,bulk}}\right)^2
\Biggl[\frac{\chi'_{00H}}{E_{res}(1\hbox{S}H)-\hbar\omega-i{\mit\Gamma}_H}\nonumber\\
&+&\frac{\chi'_{10H}}{E_{res}(2\hbox{S}H)-\hbar\omega-i{\mit\Gamma}_H}+
\frac{\chi'_{01H}}{E_{res}(1\hbox{P}H)-\hbar\omega-i{\mit\Gamma}_H}\Biggr]\\
&&+
\frac{1}{3}A_{ehL}\epsilon_1\Delta_{LT}\left(\frac{\mu_{\parl\,L}}{\mu_{\parl\,bulk}}\right)^2\nonumber\\
&&\times\left[\frac{\chi'_{00L}}{E_{res}(1\hbox{S}L)
-\hbar\omega-i{\mit\Gamma}_L}
+\frac{\chi'_{01L}}{E_{res}(1\hbox{P}L)-\hbar\omega-i{\mit\Gamma}_L}\right],\nonumber\earr
where  \bearr\label{def2}
\chi'_{00H}&=&\pi\frac{16}{(1+2\rho_{0H})^4},\nonumber\\
\chi'_{100}&=&0.6\,\pi\;\left(\frac{3}{3+2\rho_{0H}}\right)^6(1-2\rho_0)^2,\nonumber\\
\chi'_{01H}&=&1.6\,\left(\frac{3}{3+2\rho_{0H}}\right)^6,\\
\chi'_{00L}&=&\pi\frac{16}{(1+2\rho_{0L})^4},\nonumber\\
\chi'_{01L}&=&1.6\,\left(\frac{3}{3+2\rho_{0L}}\right)^6.
\nonumber\earr The quantities $A_{ehv}$ are related to the overlap
integrals
\begin{eqnarray}
&&A_{ehv}=\frac{a^*}{L}\left|\langle
\psi_e(z_e)\delta(z_e-z_h)\psi_{hv}(z_h)\rangle\right|^2\nonumber\\
&&=\left(\frac{a^*_{bulk}}{L}\right)a^*_{ez}a^*_{hzv}.
\end{eqnarray}
\\
The quantities $A_{ehv}$ determine the relation between the maxima
of the heavy-hole and light-hole resonances
\bearr\label{proporcja} &&\frac{\hbox{\small max H}}{\hbox{\small
max L}}=3\frac{A_{ehH}{\mit \Gamma}_L}{A_{ehL}{\mit
\Gamma}_H}\left(\frac{\mu_{\parl\,H}}{\mu_{\parl\,L}}\right)^2\left(\frac{1+2\rho_{0L}}{1+2\rho_{0H}}\right)^4\,\frac{m_{hzL}}{m_{hzH}}\nonumber\\
&&=3\frac{{\mit \Gamma}_L}{{\mit
\Gamma}_H}\left(\frac{\mu_{\parl\,H}}{\mu_{\parl\,L}}\right)^2\left(\frac{1+2\rho_{0L}}{1+2\rho_{0H}}\right)^4\,\frac{\gamma_1-2\gamma_2}{\gamma_1+2\gamma_2}.\earr

\begin{figure}[ht!]
\centering
\includegraphics[width=.9\linewidth]{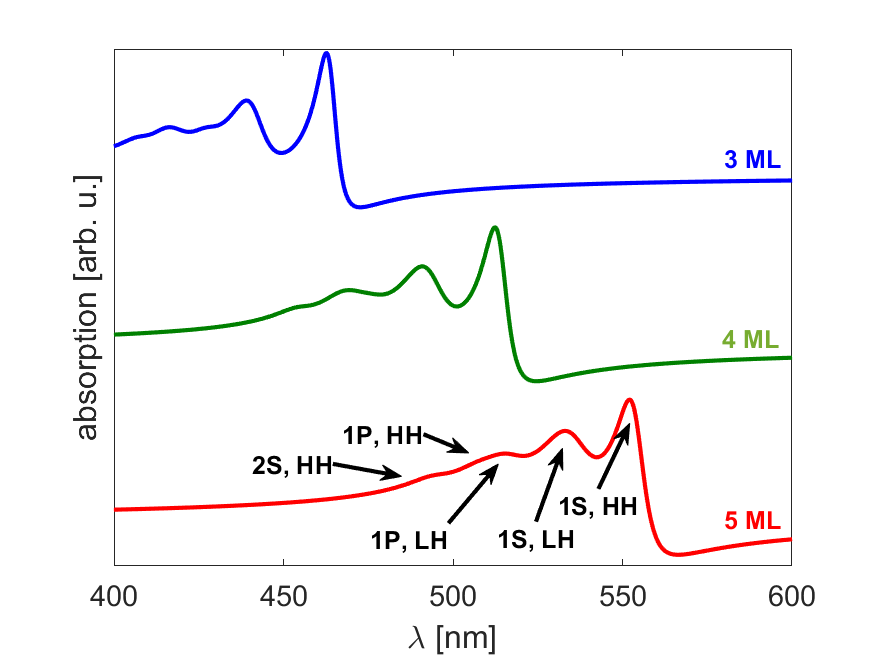}
\caption{Calculated absorption spectra for 3,4,5 monolayers platelet. Individual excitonic states are indicated by arrows.}\label{fig:abs}
\end{figure}

\begin{figure}[ht!]
\centering
\includegraphics[width=.9\linewidth]{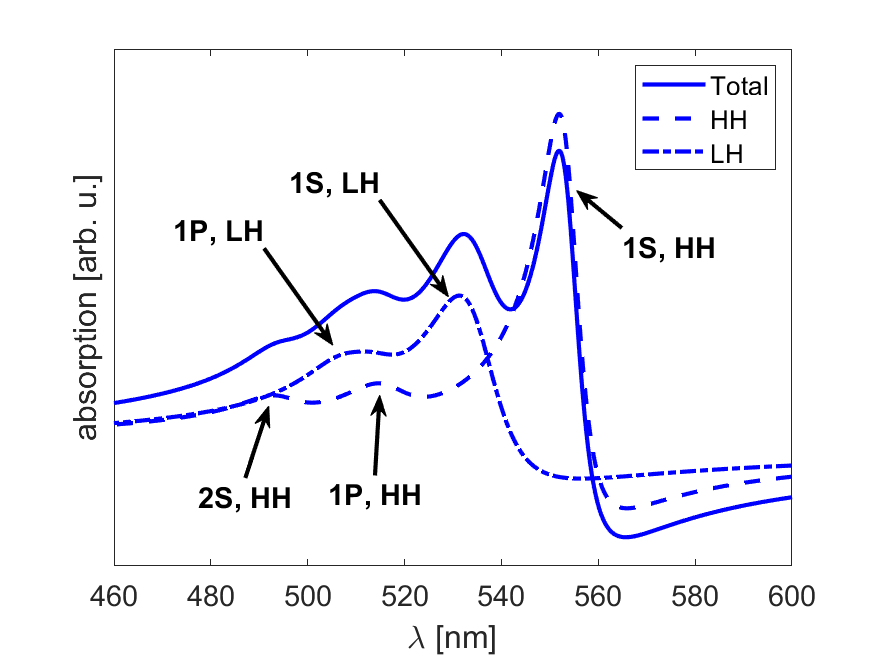}
\caption{Calculated absorption spectra for 5 monolayers platelet, divided into ligh- and heavy hole contribution. Individual excitonic states are indicated by arrows.}\label{fig:abs2}
\end{figure}

The calculated absorption coefficient $A=1-R-T$ is shown on Fig. \ref{fig:abs}.
We obtain an excellent agreement with measured spectra in literature \cite{Brumberg,Zelewski,Yu,Bagda,Shornikova,Smirnov,Ji}. The calculated binding energies and resonance energies corresponding to individual peaks in the absorption spectrum are summarized in Table \ref{tab_E}. For a particular example, for $L=4$ monolayers, $1S$ heavy hole peak is located at 2410 meV, which corresponds to $\lambda=514$ nm. In comparison, the measured peak positions for this thickness are 513 nm \cite{Brumberg}, 493 nm \cite{Shornikova}, 512 nm \cite{Ji,Yu}. The dominant feature of the spectrum are two main peaks corresponding to heavy- and light hole $1S$ exciton, located approximately 30-40 nm apart, consistently with Refs. \cite{Brumberg, Shornikova, Zelewski, Bagda, Smirnov}. The light hole peak is usually wider and in some experiments, it seems to have a complex structure \cite{Zelewski}. We can attribute this to an overlap of several weaker maxima corresponding to $1P$ and $2S$ states of light and heavy holes (see arrows on Fig. \ref{fig:abs}). To further clarify the structure of the spectrum, Fig. \ref{fig:abs2} depicts the individual contributions of light and heavy hole excitons to the spectrum of 5 monolayer platelet.
Noticeably, $1P$ states of light and heavy holes overlap almost completely, making them impossible to distinguish in the total spectrum. Note that the linewidths $\Gamma_v$ of individual state absorption lines were fitted to match the experimental spectra, in particular \cite{Zelewski}, where some of the fine features of the spectrum discussed here can be seen in the measurements.

\begin{figure}[ht!]
\centering
\includegraphics[width=.9\linewidth]{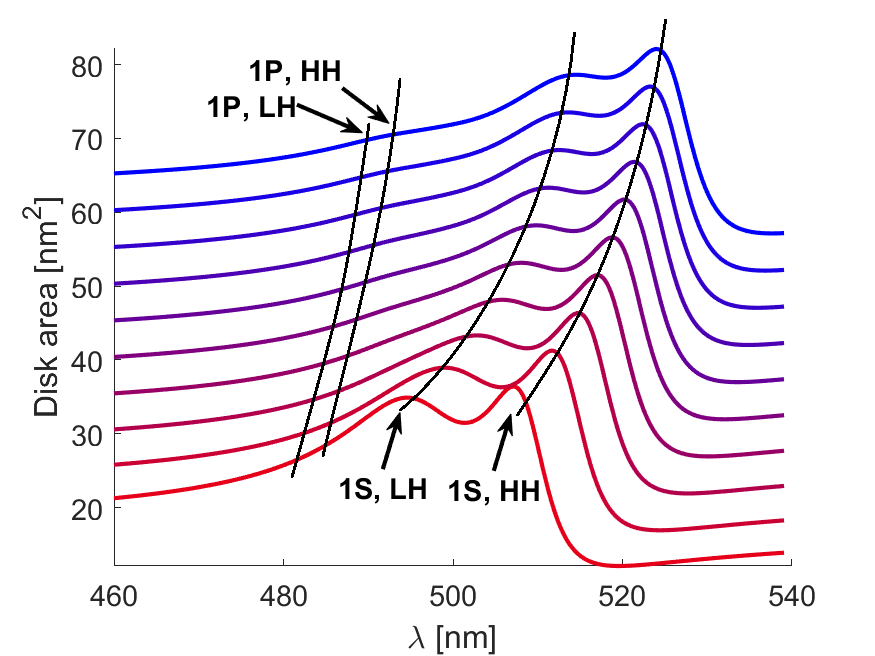}
\caption{Calculated absorption spectra for 4 monolayer disks with various areas. Individual excitonic states are indicated by arrows.}\label{fig:abs3}
\end{figure}

The absorption spectrum of a disk is also calculated from Eq. (\ref{chi1}), but with the appropriate resonance energies given by denominator of Eq. (\ref{cjdisk}). The results are shown on Fig. \ref{fig:abs3}.
In the limit of large area, the peaks of absorption spectrum asymptotically approach the energies measured by \cite{Brumberg}, where square plates with area on the order of 100 nm$^2$ were considered. In general, the wavelength of exciton peaks increases with the disk area, with the largest change occurring for $1S$ state.

\begin{table}[ht!]
\centering
\begin{tabular}{llll}
\hline
Parameter & 3ML & 4ML & 5ML \\ \hline
L & 1 & 1.33 & 1.67 \\
$E_b(1SH)$ & 313.75 & 258.00 & 220.82 \\
$E_b(2SH)$ & 39.22 & 32.86 & 28.57 \\
$E_b(1PH)$ & 95.80 & 80.11 & 69.54 \\
$E_b(1SL)$ & 336.88 & 255.48 & 204.86 \\
$E_b(1PL)$ & 179.57 & 147.80 & 126.54 \\
$E_{conf,H}$ & 1170 & 839 & 667 \\
$E_{conf,L}$ & 1528 & 1060 & 819 \\ \hline

$E_{res}(1SH)$ & 2670 & 2410 & 2237 \\
$E_{res}(2SH)$ & 2954 & 2649 & 2439 \\
$E_{res}(1PH)$ & 2896 & 2599 & 2396 \\
$E_{res}(1SL)$ & 2808 & 2511 & 2312 \\
$E_{res}(1PL)$ & 2974 & 2632 & 2399 \\
$S-P$ split & 218 & 178 & 153 \\
$S-P$ split \cite{Shornikova} & 214 & 181 & 154 \\\hline
\end{tabular}
\caption{Binding energies and resonance energies of a platelet [meV], electron mass from \cite{Benchamekh}}\label{tab_E}
\end{table}

\section{Conclusions}
In this paper we have discussed some remarkable optical properties of CdSe monolayers systems.
Atomically thin CdSe NPLs have unique physical properties which could be valuable for a broad range of applications \cite{Yu}, \cite{Bai}. Strong light-matter interaction and atomically thin volume are advantages for 2D semiconductors which make them easy tunable, as the optical properties can be controlled using multiple modulation methods.  The remarkable  thinness of these materials also provides   unique opportunities for engineering the excitonic properties. For example, changing the dielectric environment of NPLs significantly reduces the exciton binding energies and the free-particle band gap. With the help of RDMA, using dielectric potential resulting from the dielectric confinement, we have derived analytical expressions for the  binding energy and absorption for   systems of NPLs and disks depending on the monolayers number. Our results have been  thoroughly discussed and compared with the available experimental data showing a fairly good agreement. This approach and results may open up a variety of possibilities to manipulate excitonic states on the nanometer scale in 2D materials in the future. 

\appendix
\section{Eigenfunctions of the dielectric confinement}\label{Appendix A}
We consider the equation
\begin{eqnarray}
&&\label{2}-\frac{\hbar^2}{2m_z}\frac{d^2}{dz^2}\psi(z)
+\frac{\gamma\,R_z^*\,a_z^*}{(L/2)-z}\psi(z)=E_z\psi(z),
\end{eqnarray}
 Using the relation \beq
\frac{2m_z}{\hbar^2}=\frac{1}{R_z^*a_z^{*2}},\eeq where $m_z,
R_z^*, a_z^*$ are taken for a given quasi particle, and
denoting, \bearr \label{notation} && \ell=\frac{L}{2a_z^*},\quad
\zeta=\frac{\ell -z}{a_z^*},\nonumber\\
&&\varepsilon=\frac{E}{R^*},\quad \kappa=2\sqrt{\epsilon},\quad
z'=-i\kappa \zeta,\earr we transform the Eq. (\ref{2}) into the
form \beq\label{3}
\frac{d^2\psi}{dz'^2}+\left[-\frac{1}{4}+\frac{\lambda}{z'}+\frac{\mu^2-1/4}{z'^2}\right]\psi=0,\eeq
where
$$\lambda=-\frac{i\gamma}{\kappa},\quad \mu=\frac{1}{2}.$$
The equation (\ref{3})
 has the form of a Whittaker equation. The solutions of
(\ref{3}) are the Whittaker functions (see, for example,
\cite{Abramowitz}) $M_{\lambda,\mu}(x), W_{\lambda,\mu}(x)$; we use the first one in the form
\bearr
&&M_{\lambda,\mu}(z')=\\
&&=x^{\mu+1/2}\,e^{-z'/2}\,M\left(\mu-\lambda+\frac{1}{2},\,2\mu+1;z'\right),\nonumber\earr
with the confluent hypergeometric function $M(a,b;x)$.

 Substituting
$\mu=1/2$ we obtain the solution of Eq. (\ref{1}) in the form
\beq\label{eigenfunction}
\psi(z')=C\,z'\,\,e^{-'/2}\,M\left(1-\lambda,2;z'\right),\eeq with
the normalization constant $C$.

The confinement condition requires  vanishing of the
eigenfunction on the platelet interfaces. The condition
$\psi(z=L/2)=0$ is automatically satisfied, as follows from the
formula (\ref{eigenfunction}). The condition $\psi(z=-L/2)=0$,
which can be written in the form \beq\label{bc}
\psi(-i\kappa\ell)=0,\eeq will be used to calculate the
confinement states energies.

As it follows from Eq. (\ref{polarization}), only the real part of
the function $Y$ contributes. Therefore we solve the Eq.
(\ref{bc}) by putting \beq\label{condition2}
\hbox{Re}\,M\left(1+\frac{i\gamma}{\kappa},2;-i\kappa\ell\right)=0.\eeq
Taking the lowest terms from the expansion (\ref{series}) and
leaving the terms linear in $\gamma$, we obtain the relation
\begin{eqnarray}\label{eigenvalue}
&&1+\frac{\gamma\ell}{2}-\frac{1}{6}\kappa^2\ell^2=0.\end{eqnarray}
Using the definitions of $\kappa$ and $\ell$, we arrive at the
eigenenergies of the dielectric confinement \bearr\label{ehz3}
&&E_{e,hz}=\frac{6\beta[1+(\gamma_{e,h}\ell_{e,h}/2)]}{m_{e,hz}L^2},\earr
which give the total confinement energy \bearr\label{totalconf}
&&E_{conf}=E_{ez}+E_{hz}\\
&&=\frac{1}{\mu_z}\frac{6\beta}{L^2}+\frac{3\beta\gamma_{e}\,\ell_e}{m_{ez}L^2}
+\frac{3\beta\gamma_h\,\ell_h}{m_{hz}L^2},\nonumber\earr
 with $\beta$
\beq \beta= a_B^{*2}\times R^*_B= 38\,\hbox{nm}^2\,\hbox{meV},\eeq
where $[L]=\,\hbox{nm}$.
 Using the value of
 $\beta$, and the definitions
 $$\ell_{e,h}=\frac{L}{2a^*_{ze,h}},\quad
 a^*_{ze,h}=\frac{1}{m_{ze,h}}\epsilon_b\,a^*_B,$$
 we obtain the confinement energy
 \bearr\label{conf24}
 E_{conf v}(L)&=&\frac{1}{\mu_{zv} L^2}{228}[\hbox{nm}^2\hbox{meV}]\nonumber\\
 &+&\frac{\gamma}{L}\,359[\hbox{nm}\,\hbox{meV}],\earr
where
\begin{equation}\label{gamma}
\gamma=\gamma_e+\gamma_h,\quad v=H,L\end{equation} with $E$ given
in meV and $L$ in nm. The factor 359 is obtained from the relation
\beq
\frac{3\beta}{L}\left(\frac{\ell_e}{m_{ez}}+\frac{\ell_h}{m_{hz}}\right)=\frac{3\beta}{\epsilon_1\,a^*_B},\eeq
when using $\epsilon_1=6$.

Using Eq. (\ref{eigenfunction}), where we take the lowest three
terms and (\ref{eigenvalue}), we obtain the
normalized eigenfunctions for the dielectric confinement in the
form
\begin{eqnarray}\label{eigendiel}
&&\psi_{e,h}(z_{e,h})= C(-i\kappa_{e,h}
z_{e,h})\exp\left[\left(\frac{i\kappa_{e,h}z_{e,h}}{2}\right)\right]\nonumber\\
&&\times\hbox{Re}\,\left[\,M\left(1+\frac{i\gamma}{\kappa},2;-i\kappa
z_{e,h}\right)\right]\nonumber\\
&&=C(-i\kappa_{e,h}
z_{e,h})\exp\left[\left(\frac{i\kappa_{e,h}z_{e,h}}{2}\right)\right]\nonumber\\
&&\times\sqrt{\frac{13.12}{{\mathcal
L}_{e,h}}}\frac{\zeta_{e,h}}{{\mathcal
L}_{e,h}}\left(1-\frac{\zeta^2_{e,h}}{{\mathcal
L}^2_{e,h}}\right),\end{eqnarray} where
$$\zeta_{e,h}=\frac{z_{e,h}}{a^*_{e,hzv}},\quad {\mathcal
L}_{e,hv}=\frac{L}{a^*_{e,hzv}}.$$

\section{Estimation of effective masses}
\begin{figure}[ht!]
\centering
\includegraphics[width=.9\linewidth]{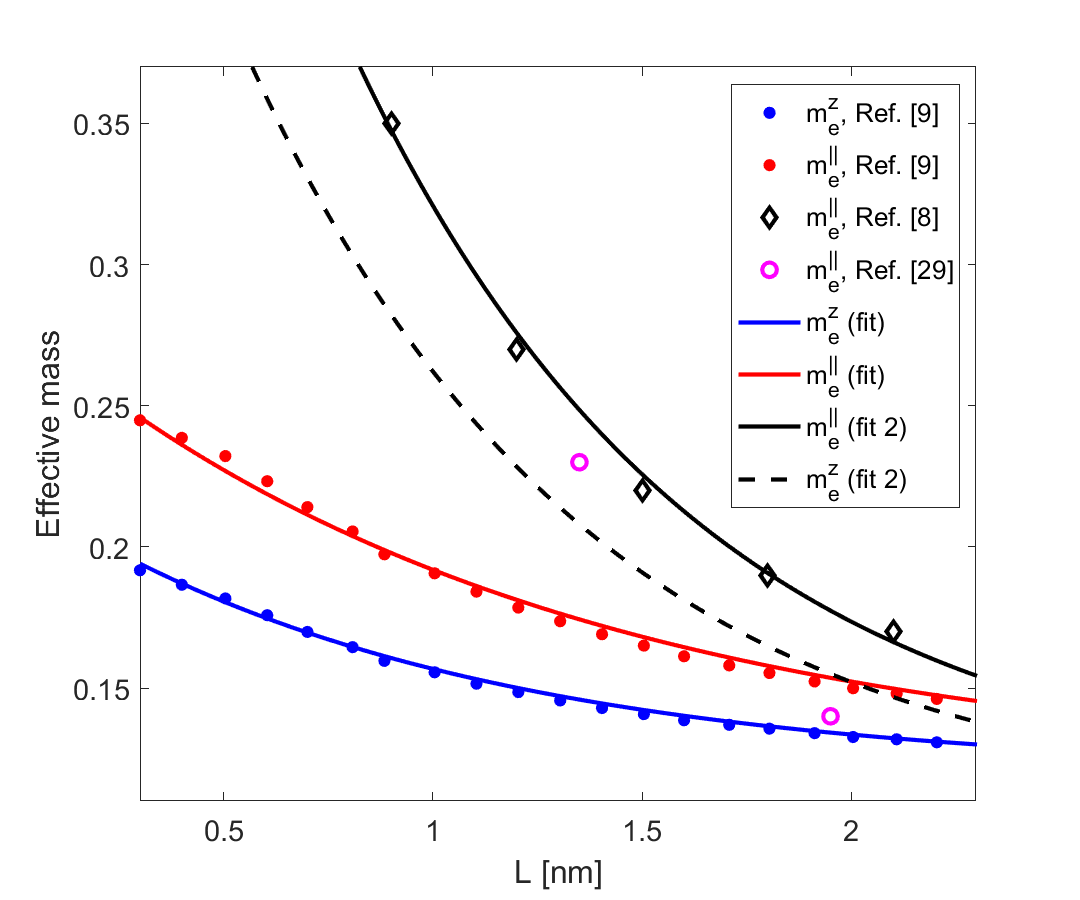}
\caption{Electron mass in CdSe nanoplatelet, fitted to the data from Refs. \cite{Shornikova,Benchamekh,Yeo} .}\label{fig:1}
\end{figure}

There are several literature sources regarding the electron effective  mass in CdSe nanoplatelets. One set of masses, obtained in \cite{Shornikova} from numerical calculations, can be described with exponential fits
\begin{eqnarray}\label{fits_me}
m_{ez} &=& 0.12 + 0.1 \exp(-L),\nonumber\\
m_{e\parallel} &=& 0.12 + 0.16 \exp(-0.8L),\nonumber\\
\end{eqnarray}
that are shown on Fig. \ref{fig:1}. 
However, we note that these mass calculations are based on an assumption of constant hole masses ($m_{hzH}=0.9$, $m_{h\parallel H}=0.19$), which do not match other sources. On the other hand, in \cite{Benchamekh} an almost 40$\%$ larger mass $m_{ez}$ is reported. It can be described by equation
\begin{equation}\label{fits_me2}
m_{ez}=0.11+0.7 \exp(-1.2L).
\end{equation}
For the in-plane mass, we assume that the ratio $m_{ez}/m_{e\parallel}$ from \cite{Shornikova} is maintained, resulting in a relation
\begin{equation}\label{fits_me3}
m_{e\parallel}=0.09+0.56 \exp(-1.2L).
\end{equation}
Finally, we note that some of the difference in the masses originates from the fact that the calculations in \cite{Benchamekh} are performed for room temperature, while in \cite{Shornikova} $T=4.2$ K is used. Nevertheless, larger electron masses used in our calculations yield a better match to all spectra, including low-temperature ones.

For the heavy holes, we use a fit to the data from \cite{Benchamekh} for H1 hole, in the form
\begin{eqnarray}\label{fit_mhHp}
m_{hH}^\parallel = 0.6 \exp(-L)+0.275
\end{eqnarray}
The literature data for the mass $m_{hH}^z$ are scarce; a constant value of $m_h^z = 0.9$ is proposed in \cite{Shornikova}, which is close to the generally accepted bulk value \cite{Kashuba}; however, for the best fit of binding energy, we propose the relation
\begin{eqnarray}\label{fit_mhHz}
m_{hH}^z = 2.1 \exp(-L)+0.42.
\end{eqnarray}
The above fit approaches the above mentioned value of 0.9 in the limit of thick layer ($1.5$ nm $\approx$ 5 monolayers), which is close to the limit of the applicability of our description. This fit not only results in the binding energy consistent with experimental measurements, but also allows us to estimate the light hole masses via Luttinger parameters and calculate the absorption spectrum containing both light- and heavy hole resonances consistent with experimental data.

The Luttinger parameters are given by
\begin{eqnarray}\label{masyefektqw}
m_{hH}^z&=&\frac{m_0}{\gamma_1-2\gamma_2},\nonumber\\
m_{hH}^\parallel &=&\frac{m_0}{\gamma_1+\gamma_2},\nonumber\\
m_{hL}^z &=&\frac{m_0}{\gamma_1+2\gamma_2},\\
m_{hL}^\parallel &=&\frac{m_0}{\gamma_1-\gamma_2}.\nonumber
\end{eqnarray}
With known masses $m_{hH}^z$, $m_{hH}^\parallel$, one can calculate $\gamma_1$, $\gamma_2$ from the relations
\begin{eqnarray}\label{Lut1}
&&\gamma_1=\frac{m_{h\parallel H}+2m_{hzH}}{3m_{hzH}m_{h\parallel H}},\nonumber\\
&&\gamma_2=\frac{m_{hz H}-m_{h\parallel H}}{3m_{hzH}m_{h\parallel H}}.
\end{eqnarray}
In the limit of a thin layer, the Luttinger parameters are consistent with those calculated in \cite{Laheld} for the case of a small quantum dot, which are $\gamma_1=1.66$, $\gamma_2=0.41$. With this, we can calculate light hole masses. 

Next, we calculate the relevant reduced masses
\begin{eqnarray}
\mu_{zH,L} &=& \left(\frac{1}{m_e^z}+\frac{1}{m_{hH,L}^z}\right)^{-1},\nonumber\\
\mu_{\parallel H,L} &=& \left(\frac{1}{m_e^\parallel}+\frac{1}{m_{hH,L}^\parallel}\right)^{-1}
\end{eqnarray}
and the Rydberg energies
\begin{eqnarray}
R_{ezH,L}^* = \frac{\mu_{zH,L}}{\epsilon_1^2}13.6~eV,\nonumber\\
R_{\parallel H,L}^* = \frac{\mu_{\parallel H,L}}{\epsilon_1^2}13.6~eV
\end{eqnarray}
We assume $\epsilon_1=6$, $\epsilon_2=2$ and
\begin{equation}
\gamma = \gamma_e+\gamma_h = 2\frac{1}{4}\frac{\epsilon_1-\epsilon_2}{\epsilon_1+\epsilon_2}=\frac{1}{4}
\end{equation}

In the calculations, we find the best fit to this data when using the electron and hole mass from \cite{Benchamekh}, e.g. Eqs. (\ref{fits_me2},\ref{fits_me3}) and $m_{hzH}$ given by Eq. (\ref{fit_mhHz}). For the in-plane mass, we use Eq. (\ref{fit_mhHp}), which is a fit to the data in \cite{Benchamekh}. With such a set of parameters, we obtain a binding energy that is consistent with both \cite{Benchamekh} and \cite{Shornikova}. Importantly, such a set of parameters results in a calculated absorption spectrum that is in a good agreement with multiple experimental results \cite{Brumberg,Zelewski,Yu,Bagda,Shornikova,Smirnov}. The relevant parameters calculated for 3,4,5 monolayers are shown in Table \ref{tab:masy}. We assume the monolayer thickness of $1/3$ nm.
\begin{table}[ht!]
\centering
\begin{tabular}{llll}
\hline
Parameter & 3ML & 4ML & 5ML \\ \hline
L & 1 & 1.33 & 1.67 \\ 
$m_{ez}$ & 0.2567 & 0.2015 & 0.1635 \\
$m_{e\parallel}$ & 0.3208 & 0.2519 & 0.2044 \\ 
$m_{hzH}$ & 1.1925 & 0.9754 & 0.8153 \\
$m_{h\parallel H}$ & 0.4957 & 0.4337 & 0.3879 \\
$m_{hzL}$ & 0.4149 & 0.3659 & 0.3302 \\
$m_{h\parallel L}$ & 0.8121 & 0.6887 & 0.5963 \\
\hline
$\mu_{z H}$ & 0.2112 & 0.1670 & 0.1362 \\
$\mu_{\parallel H}$ & 0.1948 & 0.1593 & 0.1338 \\
$\mu_{z L}$ & 0.1586 & 0.1300 & 0.1094 \\
$\mu_{\parallel L}$ & 0.2300 & 0.1844 & 0.1522 \\
$p_H=\mu_{\parallel H}/m_{ez}$ & 0.7589 & 0.7907 & 0.8165 \\
$p_L=\mu_{\parallel L}/m_{ez}$ & 0.8960 & 0.9152 & 0.9298 \\
\hline
$R^*_{\parallel H}$ & 73.58 & 60.20 & 51.28 \\
$R^*_{\parallel L}$ & 86.88 & 69.67 & 58.39 \\
\hline
$\gamma_1$ & 1.6243 & 1.8789 & 2.1062 \\
$\gamma_2$ & 0.3929 & 0.4269 & 0.4488 \\ \hline
\end{tabular}
\caption{Masses, reduced masses, Rydberg energies, Luttinger parameters}\label{tab:masy}
\end{table}

\end{document}